# Hysteresis and ribbons in Taylor-Couette flow of a semidilute non-colloidal suspension

Changwoo Kang [1,4], Michael F. Schatz[2] and Parisa Mirbod [3 a]

[1]Department of Mechanical Engineering, Jeonbuk National University, 567 Baekje-daero, Deokjin-gu, Jeonju-si, Jeollabuk-do, 54896, Republic of Korea

[2]School of Physics, Georgia Institute of Technology, Atlanta, Georgia 30332, USA

[3]Department of Mechanical and Industrial Engineering, University of Illinois at Chicago, 842 W. Taylor Street, Chicago, IL 60607, USA

[4]Laboratory for Renewable Energy and Sector Coupling, Jeonbuk National University, 567 Baekje-daero, Deokjin-gu, Jeonju-si, Jeollabuk-do, 54896, Republic of Korea

[a] Corresponding Author: Email address for correspondence: pmirbod@uic.edu (P. Mirbod)

**Abstract**

In this study, we numerically investigate hysteretic behaviors of secondary bifurcations in the Taylor-Couette flow of a semidilute, neutrally buoyant, and noncolloidal suspension. We consider a suspension with a bulk particle volume fraction $\phi_b$ = 0.1, radius ratio $\eta$ (= $r_i/r_o$) = 0.877 and particle size $\epsilon$ (= $d/a$) = 60, where $d$ is the gap width between two cylinders, $a$ is the radius of the suspended particles and $r_i$ and $r_o$ are the radii of the inner and outer cylinders, respectively. The suspension balance model (SBM) is introduced for numerical simulations to model the dynamics of particles undergoing shear-induced particle migration with rheological constitutive laws. The suspension flow transitions from circular Couette flow (CCF) via ribbons (RIB), spiral vortex flow (SVF) and wavy spiral vortex flow (WSVF) to wavy vortex flow (WVF) with the increase of suspension Reynolds number ($Re_s$). The primary transition arises very slowly with an oscillatory critical mode and appears through a supercritical (or non-hysteretic) bifurcation. However, we find hysteretic behaviors in subsequent bifurcations (spiral vortex flow (SVF) ↔ wavy spiral vortex flow (WSVF) and WSVF ↔ wavy vortex flow (WVF)) during increasing-$Re$ and decreasing-$Re$ procedures with a rapid-step change near the transition boundaries. The WSVF and WVF states are more sustained below the transition boundaries, when the Reynolds number is rapidly decreased in stages. However, the SVF and WSVF become WSVF and WVF more sharply with increasing *Re*, respectively. To conclude this study, we also examine in detail, a standing wave of weak counter-rotating vortices that occurs as the primary instability and analyze the wave that oscillates in time but is stationary in space in the RIB state.

# Ⅰ. Introduction

The study of dispersed particle flows, also known as suspensions, and particle migration in Taylor-Couette systems has garnered significant interest due to their relevance in natural phenomena and practical applications within chemical engineering [1-6]. Specifically, the transitions in flow behavior induced by neutrally buoyant particles suspended in a viscous fluid have been investigated through experimental [2-5,7] and numerical [6,8] studies in Taylor-Couette flow (TCF), which involves a rotating inner cylinder and a stationary outer cylinder. Understanding flow transitions in suspensions within a Taylor-Couette apparatus has been the subject of extensive research. This is particularly important as the characteristics of different flow states play a crucial role in the design of efficient industrial applications, such as chemical mixers and bioreactors. The controlled mixing and aggregation conditions within suspensions in geometries akin to Taylor-Couette systems can be leveraged to enhance performance in process technology and biotechnology applications. Examples include the aggregation and flocculation of fine particles and the post-processing of bio-printed human dermal tissue [9,10]. Consequently, the presence or absence of hysteresis in flow state transitions could lead to modified flow conditions, potentially enabling either high particle aggregation or high uniformity of mixing.

Several experiments have studied transitions in suspension TCF. Majji and Morris [1] investigated transitions in a dilute suspension ($\phi_b = 0.001$) of neutrally buoyant particles within a Taylor-Couette device with radius ratio $\eta (= r_i/r_o) = 0.877$, where $\phi_b$ represents the bulk particle volume fraction and $r_i$, $r_o$ are the inner and outer cylinder radii, respectively. They observed transitions from circular Couette flow (CCF) to Taylor vortex flow (TVF) and from TVF to wavy vortex flow (WVF), akin to the transition for the flow of a pure Newtonian fluid, as the Reynolds number (*Re*) based on the inner cylinder rotation rate is increased. They also noted distinct migration behavior: particles formed a circular equilibrium region in the vortex core for TVF, while in WVF, particles remained roughly uniformly distributed in the annulus. Moreover, Majji *et al*. [2] studied the flow transitions of semi-dilute suspensions in TCF of neutrally buoyant particles with $\eta = 0.877$. They observed various flow structures at different *Re* for particle volume fractions $0 < \phi_b \leq 0.3$ as *Re* is reduced quasi-statically. In contrast to the transition in a pure fluid, additional non-axisymmetric flow states were observed between TVF and CCF. For suspensions with $0.05 < \phi_b \leq 0.15$, the transition occurred via WVF → TVF → spiral vortex flow (SVF) → ribbons (RIB) → CCF with decreasing *Re*. Wavy spiral vortex flow (WSVF) was also detected for $\phi_b = 0.2$, leading to the sequence WSVF → WVF → WSVF → TVF → SVF → RIB → CCF with decreasing $Re$. However, for $\phi_b = 0.3$, the transition sequence was reduced to WSVF → SVF → CCF with decreasing $Re$. Ramesh et al. [3] also conducted experiments on suspensions in TCF at different values of the particle volume fraction ($\phi_b \leq 0.25$) for $\eta = 0.914$, employing both ramp-up (increasing $Re$) and ramp-down (decreasing $Re$) protocols. They observed SVF and coexisting states, such as WVF+TVF and TVF+SVF, in the suspension flows during the ramp-up experiments. However, during ramp-down, the flow always transitioned following WVF → TVF → SVF → CCF. Non-axisymmetric patterns (such as SVF and coexistence states) were not detected for $\phi_b < 0.05$. In particular, the authors found that the secondary bifurcation from TVF to WVF exhibited subcritical (hysteretic) behavior in suspension flows, while for TCF of a pure Newtonian fluid this transition is supercritical (non-hysteretic) in nature. Since then, Ramesh and Alam [4] observed interpenetrating spiral vortices (ISVs) during both ramp-up and ramp-down protocols over a range of particle volume fractions $0.08 < \phi_b \leq 0.15$ in suspension TCF with a larger radius ratio ($\eta = 0.941$) and aspect ratio ($\Gamma = 16$). In contrast to previous findings, the coexisting state of TVF+SVF appeared as a primary transition in both ramp-up and ramp-down experiments. Baroudi et al. [5] experimentally investigated hysteresis effects associated with neutrally buoyant particles in a suspension TCF with $\phi_b = 0.1$ and $\eta = 0.877$. They examined the effect of nonuniform particle distributions due to particle migration on various flow transitions and final flow states relative to the case with a uniform particle distribution. The results showed that nonuniformly distributed particles significantly impact hysteretic behavior. Moazzen et al. [7] conducted both ramp-up and ramp-down experiments for TCF of non-colloidal suspensions with $\eta = 0.914$ at $\phi_b \leq 0.28$, measuring inner cylinder torques. They found SVF bifurcated from CCF for $\phi_b \geq 0.06$, leading to a CCF → SVF → TVF → WVF transition in both ramp-up/down protocols, similar to the observations of Ramesh et al. [3] and Baroudi et al. [5]. The threshold of the primary instability decreased with increasing $\phi_b$. The primary bifurcation was supercritical for small $\phi_b$, but subcritical for $\phi_b > 0.12$. However, the secondary and tertiary bifurcations were found to be subcritical for all values of $\phi_b$ explored in their study. The authors also verified that particles suspended in a fluid increased the torque required to rotate the inner cylinder.

    Kang and Mirbod [6] carried out a numerical study for noncolloidal suspensions in TCF, employing the suspension balance model (SBM) and rheological constitutive laws of a

semidilute ($\phi_b$ = 0.1) suspension in a TCF system with $\eta = 0.877$. The primary and secondary bifurcations were examined by varying the rotating angular velocity of the inner cylinder, keeping the outer cylinder at rest. Non-axisymmetric states (SVF and WSVF) were found between CCF and WVF for the suspension of larger particles, consistent with previous experiments [2,5]. Thus, the transitions occurred following the sequence of CCF → SVF → WSVF → WVF with increasing $Re$. However, hysteresis in secondary bifurcations was not investigated since the rotating velocity of the inner cylinder was maintained at a constant during the computations, i.e. the Reynolds number was not increased or decreased as in the experiments. Kang and Mirbod [8] extended their numerical studies of TCF to concentrated non-colloidal suspensions of $\phi_b$ = 0.2 & 0.3. The effect of the particle concentration on the flow of suspensions was examined. RIB emerged as the primary instability from CCF, similar to the experiments of Majji *et al*. [2] and Baroudi *et al*. [5]. RIB appeared as weak modulated counter-rotating vortices. The vortices were axisymmetric and stationary in space but oscillated in time. With increasing $Re$, the flow underwent a transition to SVF, which traveled in the axial direction with a constant speed. With additional increases in $Re$, the flow transitioned first to WSVF, then to WVF and, finally, to modulated wavy vortex flow (MWVF) for both $\phi_b$ leading to the sequence of CCF → RIB → SVF → WSVF → WVF → MWVF. The nature of bifurcations (supercritical vs subcritical) was not examined [8]. Finally, Baroudi *et al*. [11] provided an overview of experimental, theoretical, and numerical studies for TCF of neutrally buoyant non-colloidal suspensions, emphasizing the effect of finite-size particles on the flow transitions and structures.

    This study aims to explore hysteretic behaviors in the secondary bifurcations (SVF↔WSVF and WSVF↔WVF) as observed in our prior work [6] employing SBM. To understand the effect of particle distributions on the flow transitions, we either increased or decreased the suspension Reynolds number, based on the rotating angular velocity of the inner cylinder, with a rapid-step change near the boundary of transitions similar to the experiments conducted by Baroudi et al. [5], where the Reynolds number was subjected to a rapid-step change to investigate the effect of the concentration profile on the flow transition. Our model explicitly disregards inertial migration, presuming that shear-induced particle migration dominates. In contrast, the experiments involve finite-sized particles with small (but non-negligible) inertia undergoing migration due to the presence of flow curvature in rotating flow, resulting in a non-uniform particle distribution beyond the predictive capability of SBM. Despite the marked difference in

dominant physics between simulations (shear-induced migration) and experiments (inertial migration), our past [6] and current studies capture much of the transition behaviors observed in the experiments of Majji *et al.* [2] and Baroudi *et al.* [5].

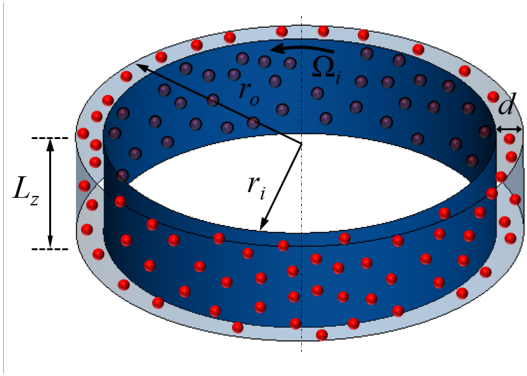

Figure 1. Schematic diagram of a suspension Taylor-Couette flow.

## II. Problem formulation

We examine a flow of neutrally buoyant, noncolloidal, rigid, spherical particles suspended in a viscous fluid of viscosity $\mu$. The suspension is contained between two coaxial cylinders where the inner cylinder rotates at an angular velocity $\Omega_i$ and the outer cylinder is stationary (Fig. 1). $r_i$ and $r_o (= r_i + d)$ are the radii of the inner and outer cylinders, respectively, where $d$ is the gap width between the two cylinders. The radius ratio ($\eta = r_i/r_o$) of cylinders is $\eta = 0.877$, and the aspect ratio ($\Gamma = L_z/d$) is $\Gamma = 4$ ($\cong 2\lambda_c/d$), where $\lambda_c$ is the critical wavelength for the flow of a pure Newtonian fluid. Relating $\Gamma$ to $\lambda_c$ is justified by the fact that Majji *et al.* [2] experimentally found states with an axial wavelength that is close to the pure fluid value. We consider a suspension of $\phi_b = 0.1$ with the particle size $\epsilon = d/a = 60$, where $a$ is the radius of the suspended particles. Similar to our previous work [6], herein it is assumed that $Re_p \ll 1$, where $Re_p = \rho\dot{\gamma}a^2/\mu$ is the particle Reynolds number based on the fluid shear rate $\dot{\gamma}$ and the radius of the suspended particles $a$, and thus the fluid-particle interactions are strong and the particle inertia is negligible. We also suppose that the suspension is in the limit of the infinite Péclet number ($Pe \to \infty$), where $Pe = 6\pi\mu\dot{\gamma}a^3/kT$ is the Péclet number defined with the thermal energy $kT$. We define the suspension Reynolds number as $Re_s$ ($= \rho r_i \Omega_i d/\mu_s(\phi_b)$) where $\mu_s(\phi_b)$ is the effective shear viscosity of the bulk suspension and the effective viscosity of the suspension is defined by Krieger's empirical correlation $\mu_s(\phi) = \mu(1 - \phi/\phi_m)^{-1.82}$ with $\phi_m = 0.68$ [12]; $Re_s$ is varied to examine transition behaviors by changing the angular velocity $\Omega_i$. For increasing-*Re* and decreasing-*Re* protocols, we use flow and concentration fields produced in the simulations of Kang and Mirbod [6] as initial conditions, which are circular Couette flow for the flow field and a uniform particle volume fraction for the concentration field. The Reynolds number increases or decreases with a step change during the protocols; after the step change, $Re_s$ is held constant until the flow becomes a fully-developed state.

The suspension balance model (SBM) [13,14] has been used to model the suspension flow. The balance equations for the suspension and the conservation equation for the particle mass are given by [6,14-16]

$$\nabla \cdot \langle \mathbf{u} \rangle = 0 \qquad (1)$$

$$\frac{D\langle \rho \mathbf{u} \rangle}{Dt} = \langle \mathbf{b} \rangle + \nabla \cdot \langle \mathbf{\Sigma} \rangle \qquad (2)$$

$$\frac{\partial \phi}{\partial t} + \langle \mathbf{u} \rangle \cdot \nabla \phi = -\nabla \cdot \mathbf{j}, \quad \mathbf{j} = \frac{2a^2}{9\mu} f(\phi) \nabla \cdot \langle \mathbf{\Sigma} \rangle_p. \qquad (3)$$

where the material derivative is $D/Dt = \partial/\partial t + \langle \mathbf{u} \rangle \cdot \nabla$ [13,16] and $\langle \ \rangle$ denotes a suspension average. Here, $\langle \mathbf{u} \rangle$, $\langle \mathbf{b} \rangle$ and $\langle \mathbf{\Sigma} \rangle$ are the suspension-averaged velocity, the average body force acting on the suspension and the average suspension stress, respectively [13,14]. In Eq. (3), $\mathbf{j}$ represents the particle migration flux, and $f(\phi)$ is the sedimentation hindrance function defined as $f(\phi) = (1 - \phi/\phi_m)(1 - \phi)^{\alpha-1}$ with $\alpha = 4$ [17]. $\phi$ is the particle volume fraction, $\phi_m = 0.68$ is the maximum packing particle volume fraction [12], and $\langle \mathbf{\Sigma} \rangle_p$ is the average particle stress [6,16].

The suspension $\langle \mathbf{\Sigma} \rangle$ and particle $\langle \mathbf{\Sigma} \rangle_p$ stresses are defined as [13,14,16]

$$\langle \mathbf{\Sigma} \rangle = -\langle p \rangle_f \mathbf{I} + 2\mu \langle \mathbf{S} \rangle + \langle \mathbf{\Sigma} \rangle_p \qquad (4)$$

$$\langle \mathbf{\Sigma} \rangle_p = -\mu_n(\phi)\dot{\gamma}\mathbf{Q} + 2\mu_p(\phi)\langle \mathbf{S} \rangle \qquad (5)$$

where $\langle p \rangle_f$ is the average pressure in the fluid, $\mathbf{I}$ is the identity tensor, and $\langle \mathbf{S} \rangle$ is the bulk suspension rate of strain given by $\mathbf{S} = [\nabla \mathbf{u} + (\nabla \mathbf{u})^T]/2$. In Eq. (5), the first term states the particle contribution to the normal stress, $\dot{\gamma} = (2\mathbf{S}:\mathbf{S})^{1/2}$ is the local shear rate, $\mu_n(\phi)$ is the "normal stress viscosity" that can be represented as $\mu_n(\phi) = K_n \mu (\phi/\phi_m)^2 (1 - \phi/\phi_m)^{-2}$, where $K_n = 0.75$ [16,17], and $\mu_p(\phi)$ is the particle contribution to the shear viscosity. The constant tensor $\mathbf{Q}$ describes the anisotropy of the normal stresses as

$$\mathbf{Q} = \begin{pmatrix} \lambda_1 & 0 & 0 \\ 0 & \lambda_2 & 0 \\ 0 & 0 & \lambda_3 \end{pmatrix} \tag{6}$$

We chose the values of $\lambda_1 = \lambda_2 = 1$ and $\lambda_3 = 0.8$ in this study because these values both guaranteed the computational stability of our simulations and demonstrated good agreement when compared with previous experimental studies (e.g., Fig. 2 of Kang and Mirbod [6]). Additionally, we found that by applying different values $\lambda_i$ explored in other works [17, 19], the results were only weakly affected. The second term of Eq. (5) indicates the particle contribution to the shear stress, and it can be combined with the shear stress of the fluid $2\mu\langle\mathbf{S}\rangle$. Therefore, the suspension stress can be expressed as

$$\langle\mathbf{\Sigma}\rangle = -\langle p \rangle_f \mathbf{I} - \mu_n(\phi)\dot{\gamma}\mathbf{Q} + 2\mu_s(\phi)\langle\mathbf{S}\rangle. \tag{7}$$

where $\mu_s(\phi) = \mu + \mu_p(\phi)$ [13,14,16]. More details on the balance equations and the constitutive laws related to the Taylor-Couette flow can be found in Kang and Mirbod [6]. The suspension average notation $\langle\ \rangle$ is omitted for simplicity hereafter.

The balance equations for the suspension and particles (1) ~ (3) were discretized using a finite volume method. A second-order central difference and QUICK (quadratic upstream interpolation for convective kinematics) schemes were used for the spatial discretization of derivatives. For time advancement, a third-order Runge–Kutta scheme and a second-order Crank-Nicolson method were employed. A fractional step method was used for time integration, and the Poisson equation resulting from the second stage of the fractional step method was solved by a fast Fourier transform (FFT).

For the boundary conditions, the no-slip condition was applied on the walls of cylinders ($\mathbf{u} = r_i\Omega_i$ at $r = r_i$, $\mathbf{u} = 0$ at $r = r_o$), and the migration flux of particles was set to zero at the walls ($\boldsymbol{j}\cdot\mathbf{n} = 0$ at $r = r_i, r_o$). The flow and particle concentrations were assumed to be periodic in the axial direction ($z$). More computational details were reported in Kang and Mirbod [6]. For non-dimensionalization, we have chosen the gap width $d$ as the scale for length, the rotating velocity of the inner cylinder $r_i\Omega_i$ as the scale for velocity, and $\rho d^2/\mu$ as the time scale. Therefore, it should be noted that all properties displayed in all plots are dimensionless. In addition, for convenience, we employ a reduced radial coordinate denoted as $x = (r - r_i)/d \in [0, 1]$.

## Ⅲ. Results and discussion

### *A. Hysteresis*

Hysteretic (or subcritical) behaviors in the flow transitions have been observed in previous suspension TCF experiments [2-5]. Majji *et al*. [2] experimentally compared the flow transitions and structures during increasing-*Re* and decreasing-*Re* ramp experiments in a quasi-steady manner. They found that the flow transitions are qualitatively the same in both cases but quantitatively different. The transitions during the increasing-*Re* ramp occurred at higher *Re* against during decreasing-*Re* experiments, showing distinct hysteresis. Ramesh *et al*. [3] also experimentally identified the hysteretic behaviors in the secondary bifurcation TVF↔WTV for semidilute suspensions during quasi-steady ramp-up and ramp-down experiments. The transitions between TVF and WVF appeared at higher *Re* during the ramp-up experiments. Moreover, coexisting states of stationary (TVF) and traveling (WTF or SVF) waves (i.e., TVF+WVF and TVF+SVF states) were found only for the ramp-up protocols. Baroudi *et al*. [5] also experimentally revealed that concentration variations exhibit hysteresis in the TVF regime. The particle distribution following inertial migration stabilized the TVF near the TVF↔WVF and TVF↔SVF transition boundaries. Lately, Moazzen *et al*. [7] observed a weakly hysteretic behavior for primary bifurcations in the flow of semi-dilute and concentrated suspensions, while secondary bifurcations were hysteretic for all concentrations showing a difference in transition Reynolds numbers between ramp-up and ramp-down experiments. We, therefore, examine numerically the existence of hysteresis in the secondary bifurcations, i.e. SVF↔WSVF and WSVF↔WVF, observed in Kang and Mirbod [6].

First, we consider the transition between SVF and WSVF (see Fig. 14 of Kang and Mirbod [6]). Beginning from fully developed flow states near the transition boundary, the Reynolds number ($Re_s$) is suddenly increased or decreased by a step change acceleration or deceleration, respectively, of the inner cylinder's rotation. Figure 2(a) displays the time evolution of the flow when $Re_s$ is stepped from $Re_s$ = 135 (i.e., the SVF regime) to $Re_s$ = 140. As seen in the plot, the flow gradually evolves to WSVF with time. This transition shows agreement with the map of flow patterns presented in Kang and Mirbod [6]. Note that the map was determined by time-dependent simulations with the initial profile of uniform particle distribution. However, when the Reynolds number is decreased to $Re_s$ = 135 again after the flow reaches to the saturated state, the WSVF state is sustained (Fig. 2(b)).

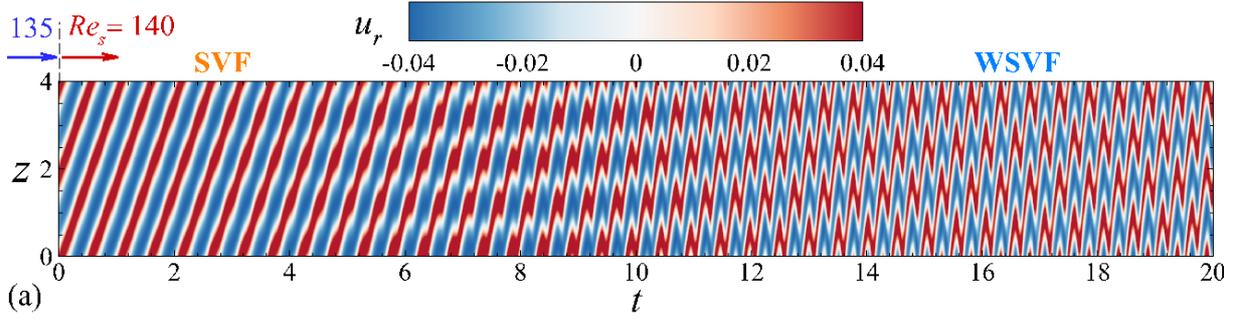

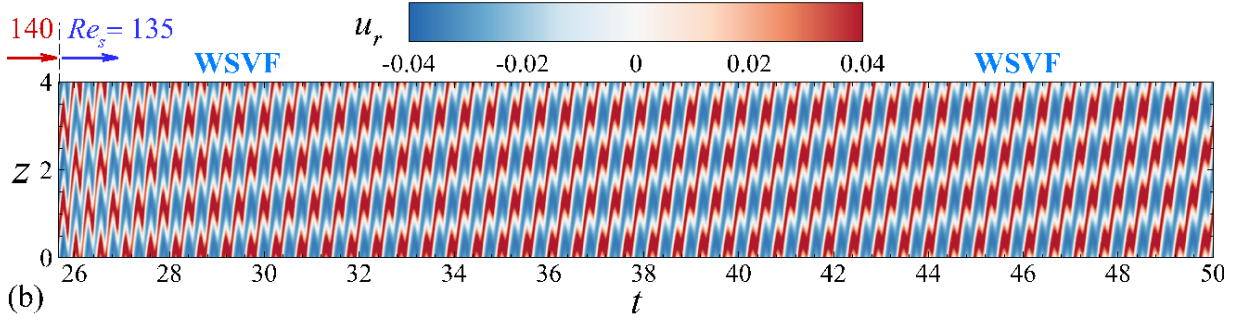

Figure 2. Space-time diagrams of radial velocity at the mid-gap ($x = 0.5$) and a given $\varphi$ for $\epsilon = d/a = 60$ and $\phi_b = 0.1$. (a) When $Re_s$ is suddenly increased from $Re_s = 135$ (SVF) to $Re_s = 140$ a transition from SVF to WSVF occurs. (b) However, when $Re_s$ is suddenly decreased to $Re_s = 135$, the WSVF state persists. Here, the axial length $z$ and time $t$ were non-dimensionalized by the gap width $d$ and $\rho d^2/\mu$, respectively.

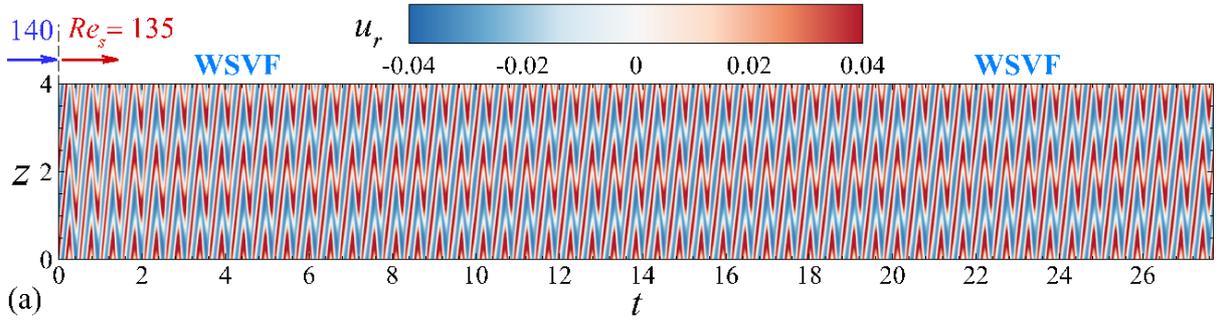

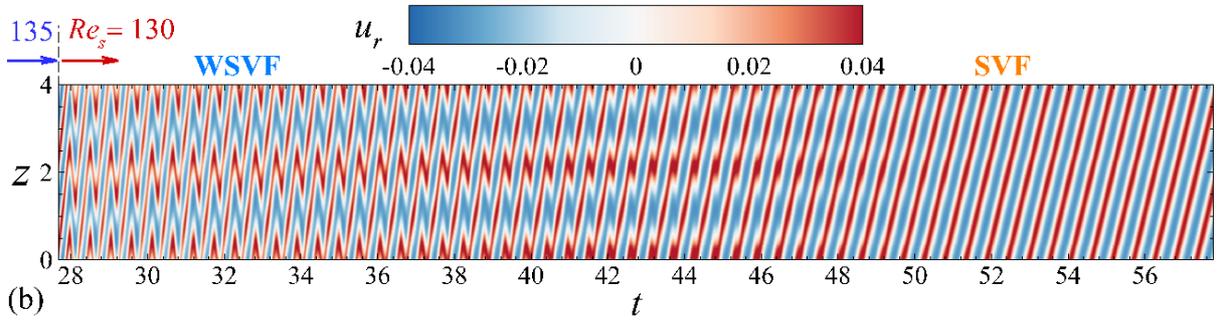

Figure 3. Space-time diagrams of radial velocity at the mid-gap ($x = 0.5$) and a given $\varphi$ for $\epsilon = 60$ and $\phi_b = 0.1$ when $Re_s$ is suddenly decreased near the boundary between SVF and WSVF; (a) from $Re_s = 140$ (WSVF) to $Re_s = 135$, (b) from $Re_s = 135$ (the end of (a)) to $Re_s = 130$.

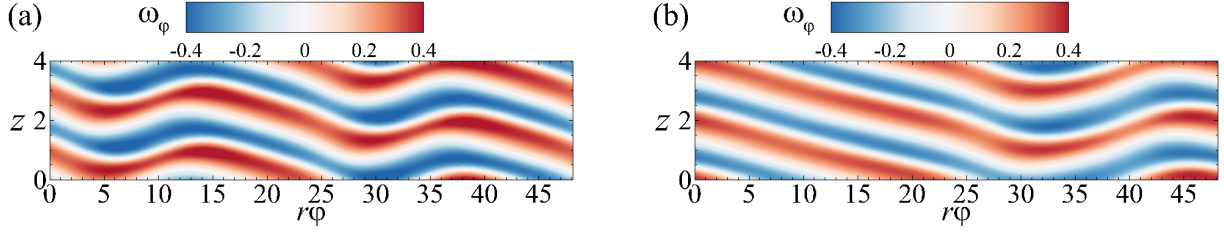

Figure 4. Instantaneous contours of azimuthal vorticity ($\omega_\varphi$) versus $r\varphi$, where $r\varphi$ indicates the nondimensionalized length in the azimuthal direction at the central surface ($x = 0.5$), (a) at the end of Fig. 2(a) and (b) at the end of Fig. 3(a).

On the other hand, the suspension flow continuously maintains the pattern (i.e., WSVF) when $Re_s$ is suddenly reduced from the WSVF state ($Re_s = 140$) to $Re_s = 135$ (Fig. 3(a)). It should be noted again that the WSVF at the initial stage of Fig. 3(a) was established in Kang and Mirbod [6] as mentioned above. Here, the patterns of the time evolution are dissimilar between the WSVF states of Fig. 2 and Fig. 3. This is because the azimuthal wavenumber of the wavy vortices is different as revealed in Fig. 4. In addition, the particle distributions represent the same feature verified in the previous study [6] for WSVF. Particles are more collected in the non-wavy parts of vortices displaying distinct bands of higher particle concentration. However, as $Re_s$ is further decreased to $Re_s = 130$ after the flow reached the equilibrium state in Fig. 3(a), the flow progressively transitions to SVF as presented in Fig. 3(b). Accordingly, the bifurcation SVF↔WSVF exhibits clear hysteresis.

We also examined the transition from WSVF to WVF (see Fig. 14 of Kang and Mirbod [6]). In the WSVF regime ($Re_s = 160$), the Reynolds number is increased to $Re_s = 165$ with a step change, corresponding to the WVF regime in the map plotted in Kang and Mirbod [6]. As presented in Fig. 5(a), the suspension flow maintains the WSVF state even though $Re_s$ is increased to the value of the WVF regime in the map. After the flow is fully developed, the Reynolds number is changed to $Re_s = 170$ again. However, we found that the WSVF state is sustained continuously. As $Re_s$ is further increased to $Re_s = 175$ with the same manner, the flow undergoes an obvious transition. As indicated by arrows in Fig. 5(b), the flow first exhibits a transient change in the traveling direction of WSVF is altered to the opposite direction first. To illustrate more clearly, the time intervals marked with the arrows in Fig. 5(a) are enlarged in Figs. 5(c) and 5(d). Beyond $t = 42$, the flow travels downward. However, the flow transitions to WVF after a sufficiently long time (Fig. 5(e)). At approximately $t = 51$, the transition occurs, and WVF appears in the annulus thereafter. Figures 5(f) and 5(g) clearly show the alternation of flow patterns. It should be noted that the particles are also almost uniformly dispersed in the annulus for WVF [6].

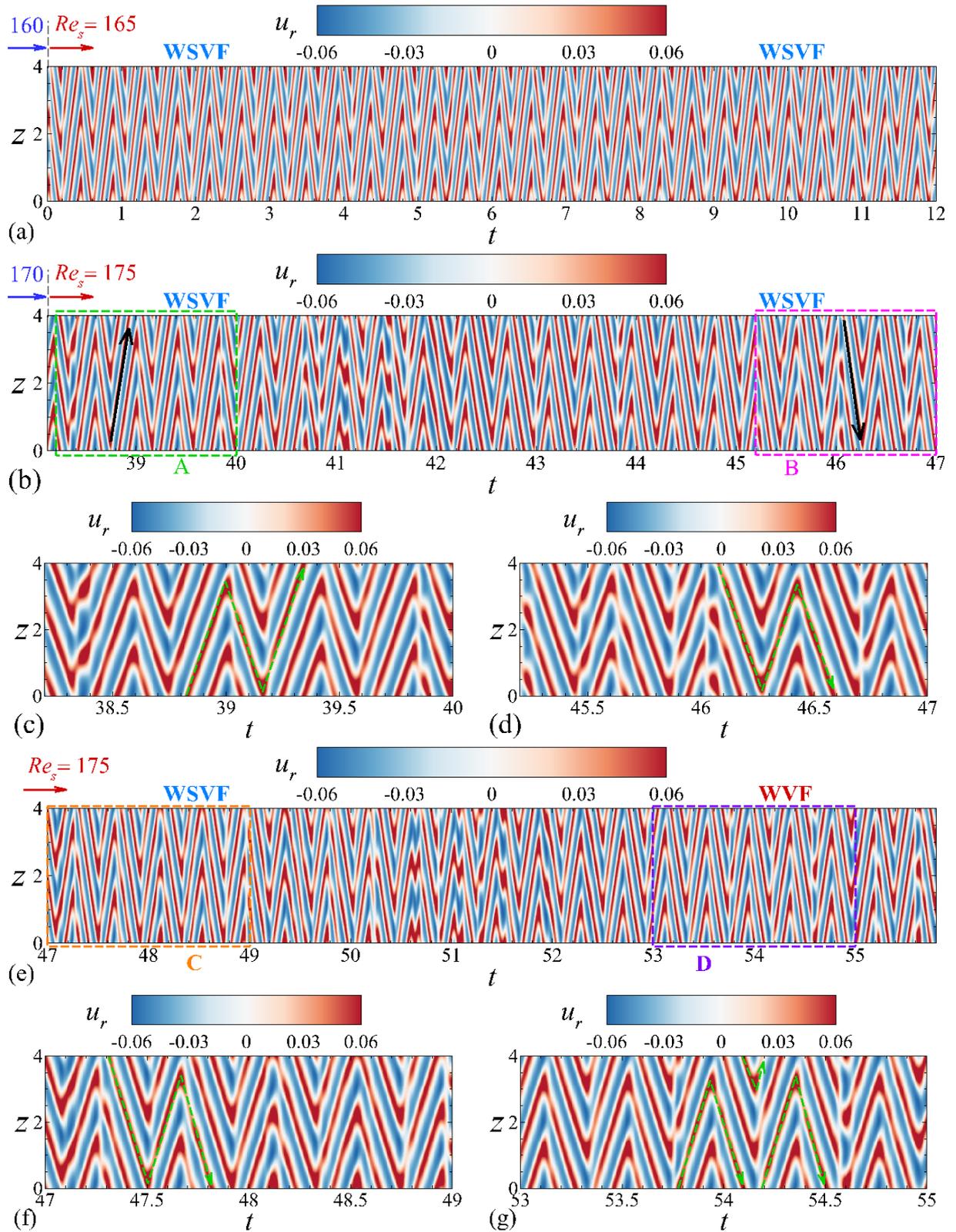

Figure 5. Space-time diagrams of radial velocity at the mid-gap ($x = 0.5$) and a given $\varphi$ for $\epsilon = 60$ and $\phi_b = 0.1$ when $Re_s$ is stepped (a) from $Re_s = 160$ (WSVF) to $Re_s = 165$ and (e) from $Re_s = 170$ to $Re_s = 175$. Zoomed-in views for (a) are shown for (c) section A and (d) section B, and zoomed-in views for (e) are shown for (f) section C and (g) section D. In (b), the arrows indicate the travel direction of WSVF.

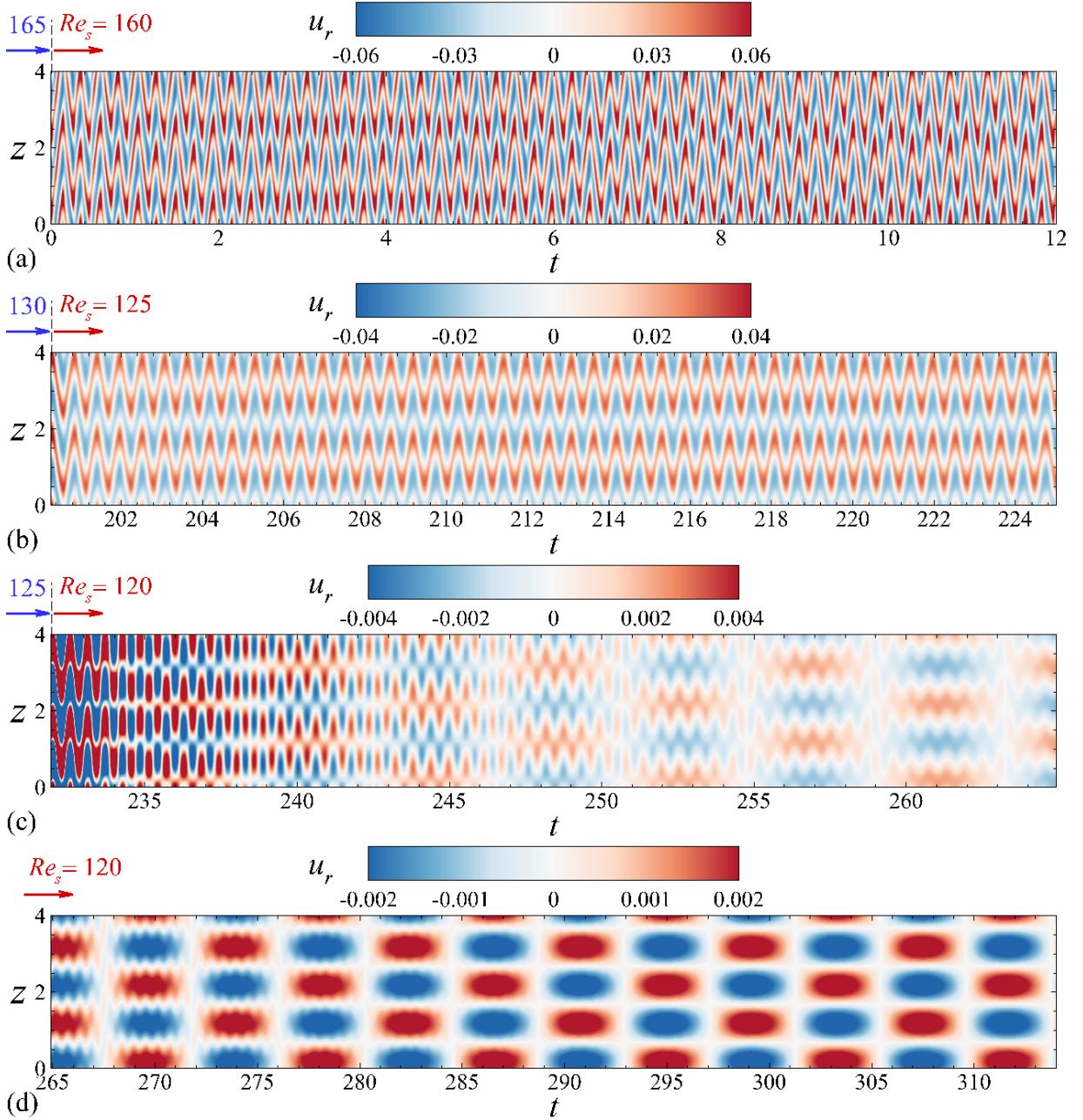

Figure 6. Space-time diagrams of radial velocity at the mid-gap ($x = 0.5$) and a given $\varphi$ for $\epsilon = 60$ and $\phi_b = 0.1$ when $Re_s$ is decreased in steps from (a) from $Re_s = 165$ (WVF) to $Re_s = 160$, (b) from $Re_s = 130$ to $Re_s = 125$, (c) from $Re_s = 125$ to $Re_s = 120$, (d) continuous plot of (c). Note that SVF and WSVF did not appear during the decreasing-$Re$ ramp.

Beginning with the flow in the WVF regime at $Re_s = 165$, we examined the flow behavior as the Reynolds number is reduced (Fig. 6(a)). A stable WVF pattern was found to persist at $Re_s = 160$, even though WSVF was found at this Reynolds number in an earlier study (see Fig. 14 of Kang and Mirbod [6]). In particular, as $Re_s$ is progressively reduced via a series of step changes (with each step representing a change of 5 in Reynolds number), the WVF state is continuously sustained until $Re_s = 125$ (Fig. 6(b)). A clear transition arises with time when the Reynolds number is further decreased to $Re_s = 120$, but not to either SVF or WSVF, as seen when ramping up in Reynolds number. Instead, the intensity of the flow steadily weakens, and the oscillation in the axial direction disappears gradually (Fig. 6(c)). Finally, the ribbon pattern (RIB), which results from time-modulated counter-rotating vortices, fills the annulus of the cylinder (Fig. 6(d)).

In summary, different types of flow transitions have been observed when $Re_s$ is suddenly increased or decreased with a step change from each flow pattern. First, the SVF↔WSVF transition occurs at different ranges of $Re_s$ with increasing-$Re_s$ and decreasing-$Re_s$ protocols (Figs. 2 and 3). In addition, dissimilar transition scenarios have been presented when $Re_s$ is increased and, subsequently, decreased near the boundary of transition WSVF↔WVF (Figs. 5 and 6). As shown in Fig. 5, with increasing $Re_s$, WSVF becomes WVF at higher $Re_s$ compared with the previous simulations [6]. However, with decreasing $Re_s$, WVF transitions to RIB instead of WSVF (Fig. 6). These differences provide clear evidence of hysteresis in the bifurcations beyond the primary transition.

### *B. Ribbons*

Ribbons (RIB) have been observed previously in experiments as the transition from CCF in quasi-steady decreasing-$Re$ experiments [2] as well as in both step-increasing and step-decreasing $Re$ protocols [5]. In the present study, we have identified the presence of the ribbons near the threshold of the instability during the step-decreasing-$Re$ ramps as shown in Fig. 6(d). While perturbations grow gradually after the random noise given at the initial state is dissipated [18], for a flow with a very small growth rate, the perturbation starts to increase after a long time and the noise vanishes very slowly. It could be then difficult to detect the growth of perturbations at the threshold of instability. We thus decided to revisit these data and examine the RIB state in detail. To find RIB flows, we carried out additional simulations for several $Re_s$ between the CCF and SVF regimes examined in the earlier study [6]. The simulations have also been conducted with the initial conditions of CCF and uniform particle concentration. To observe the growth of perturbations, the computations have been performed for a sufficiently long time well after the noise has disappeared.

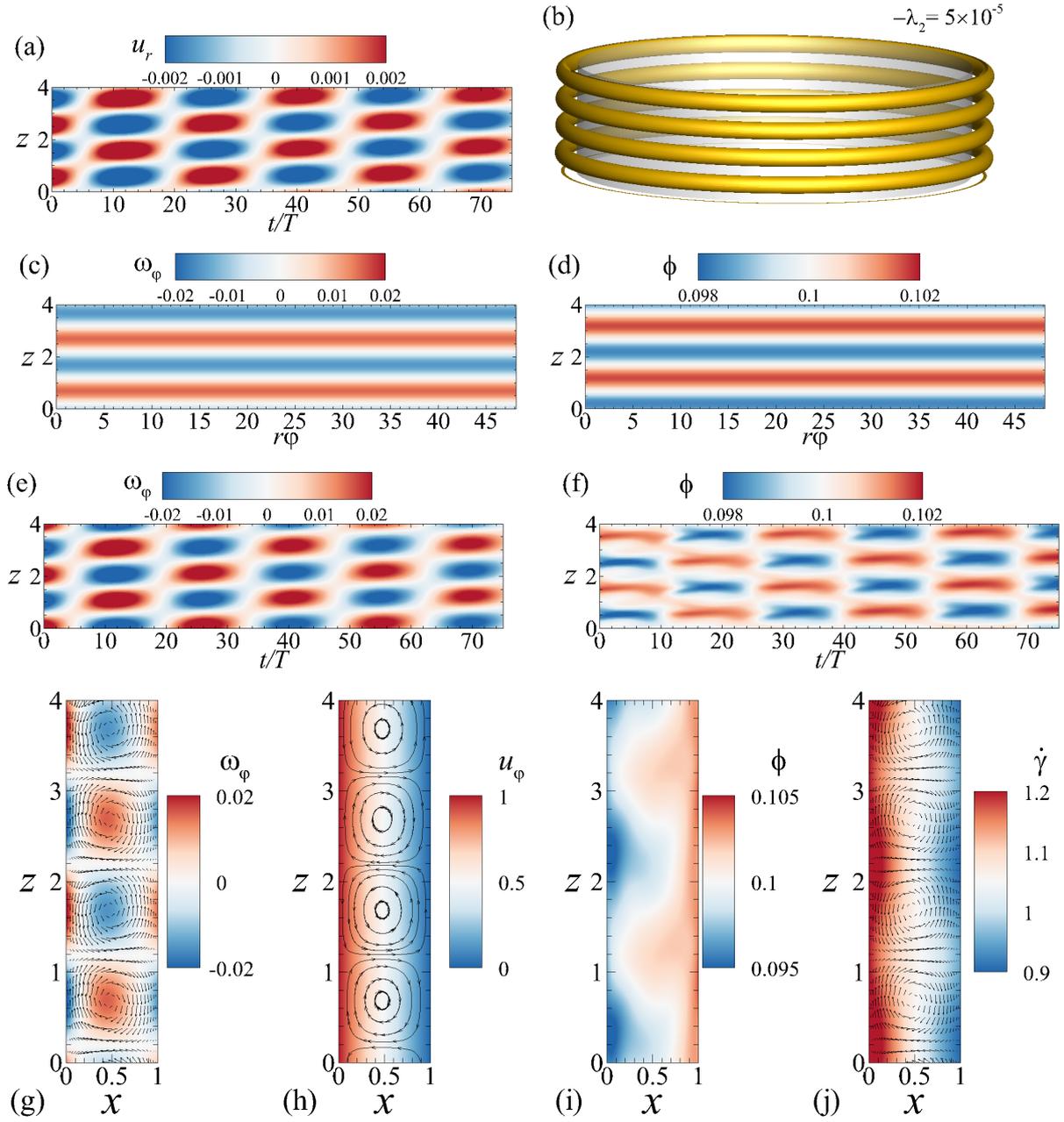

Figure 7. Flow and particle concentration fields for $Re_s = 120$, $\epsilon = 60$, and $\phi_b = 0.1$: (a) space-time diagram of radial velocity ($u_r$) at the mid-gap ($x = 0.5$) and a given $\varphi$, (b) instantaneous iso-surface of $-\lambda_2 = 5 \times 10^{-5}$ [19], instantaneous contours of (c) azimuthal vorticity ($\omega_\varphi$) and (d) particle volume fraction ($\phi$) versus $r\varphi$, where $r\varphi$ indicates the nondimensionalized length in the azimuthal direction at the central surface ($x = 0.5$), space-time diagrams of (e) azimuthal vorticity ($\omega_\varphi$) and (f) particle volume fraction ($\phi$) at the mid-gap ($x = 0.5$) and a given $\varphi$, instantaneous contours of (g) azimuthal vorticity ($\omega_\varphi$) with velocity vectors, (h) azimuthal velocity ($u_\varphi$) with streamlines, (i) particle volume fraction ($\phi$) and (j) local shear rate ($\dot\gamma$) with velocity vectors in an $r$-$z$ plane ($\varphi = \pi$). Velocity vectors were plotted for every four and two points in the radial ($r$) and axial ($z$) directions, respectively. Here, $x = (r - r_i)/d \in [0, 1]$ is a reduced radial coordinate and $T$ is the time for the inner cylinder to complete one full rotation.

We discovered the ribbon state at $Re_s = 120$ from the additional simulations examined between the CCF and SVF regimes. It should be noted that there was no ramping. In Fig. 7, the spatial and temporal characteristics of the state are clearly presented. Figure 7(a) precisely shows the space-time diagram for the RIB state. Here, $T$ is the time for the inner cylinder to complete one full rotation. The positive and negative velocity regions alternately cross up and down in time, caused by modulating counter-rotating vortices. The ribbons are standing waves in the axial direction, i.e., are the equal superposition of waves moving axially upwards and downwards [20-22]. The waves propagating axially to opposite directions with the same amplitude overlap, and then two pairs of weak axisymmetric counter-rotating vortices fill the annular cavity in the instantaneous flow field as shown in Figs. 7(b) and 7(c). To visualize the vortical structure of the RIB state, we chose a method that satisfies Galilean invariance: the $\lambda_2$ criterion, which defines vortices as regions of the flow where $\lambda_2$, the intermediate eigenvalue of the symmetric tensor $\mathbf{S}^2 + \mathbf{\Omega}^2$ is less than zero. Here, $\mathbf{S}$ and $\mathbf{\Omega}$ are the symmetric and antisymmetric parts of the vorticity gradient tensor $\nabla \mathbf{u}$ respectively. This definition accurately defines vortex cores at low Reynolds numbers [19]. Therefore, the iso-surface of negative $\lambda_2$ in Fig. 7(b) clearly expresses the vortex cores of counter-rotating vortices formed in the annulus.

Figures 7 (a-c) indicate the axial wavenumbers for RIB in the numerical simulations are similar to those found experimentally by Majji *et al*. [2] ($k_z \approx \pi$); however, RIB in simulations are found to be axisymmetric ($k_\varphi = 0$) while RIB in the experiments of Majji *et al*. [2] are non-axisymmetric ($k_\varphi \approx 2$). The ribbons also induce an axisymmetric pattern in the particle concentration field, resulting in a uniform distribution of particles in the azimuthal direction, revealing two bands of elevated particle concentration (Fig. 7(d)). The differences between the experiments and simulations in RIB state behavior can be due to differences in boundary conditions between the numerical simulations (axially periodic boundaries) and the experiments in Majji *et al*. [2] (axially no-slip/no-penetration boundaries). Non-axisymmetric RIB states can be thought of as standing waves produced by the superposition of axially counterpropagating SVF traveling waves [22]. By contrast, the axisymmetric RIB states found here are thought of as standing waves resulting from the superposition of counterpropagating TVF traveling waves. Such traveling waves are allowed by the presence of a continuous symmetry due to the axially periodic boundary conditions in our simulations; however, in experiments like those of Majji *et al*. [2], no-slip/no-penetration axial boundaries would break continuous symmetry in the axial direction and, thus, axisymmetric RIB would not be expected to be found. We speculate that non-axisymmetric RIB and/or time-independent TVF might be found if the axial boundary conditions of the simulations matched the no-slip/no-penetration boundaries of typical experiments.

Figures 7(e) and 7(f) display temporal variations of azimuthal vorticity ($\omega_\varphi$) and particle volume fraction. The particle concentration also changes with time due to the modulation of counter-rotating vortices. Accordingly, it is accepted that the two counter-rotating axisymmetric vortices are stationary in space and oscillate in time. It is compared to the SVF state, which is non-axisymmetric and traveling waves in both axial and azimuthal directions [22,23].

The instantaneous flow and concentration fields at the cross-section of the r-z plane are also displayed in Figs. 7(g) ~ (j). The weak vortices depicted in Figs. 7(g) and (h) exhibit mirror symmetry, characterized by a pair of counter-rotating vortices. These vortices contribute to the convective flow and shear gradient, potentially leading to shear-induced particle migration, as illustrated in Fig. 7(j). Consequently, particles are transported and accumulated where the shear gradient is weak as can be seen in Figs. 7(i) and 7(j). Despite the appearance of counter-rotating vortices in the annulus, the distribution of particles for the RIB state differs from that of TVF and SVF. In the axisymmetric TVF, particles migrate toward the core of the vortices, resulting in a region of higher concentration at the center of the vortices [6]. On the other hand, in the non-axisymmetric SVF, particles are concentrated only in the core of clockwise-rotating vortices [6]. These results are different from Majji $et\ al.$ [2]. As stated before, this discrepancy could be attributed to the limitations of the SBM, where lift forces due to both inertia and particle rotation are neglected, and only the normal force due to anisotropic particle pressure distribution is considered. As displayed in Fig. 7(i), particles are more accumulated near the outer cylinder ($x = 1$) for RIB, resembling the distribution in CCF [6]. The particle concentrations at the outward flow between two counter-rotating vortices are higher than those at the inward flow. However, for CCF, the particle concentration distribution is almost linear over the gap, as can be seen in Fig. 3(a) of Kang and Mirbod [6]. In RIB, the shear gradient in the radial direction appears to be greater than that in the vortices, while it is stronger near the core of vortices in TVF and SVF. Accordingly, the region of higher concentration is found where the suspension flows toward the outer wall because the outward flow reduces the shear gradient in the radial direction and thus it restricts the shear-induced migration. Indeed, in the experiments of a wide-gap Couette flow at $0.1 \leq \phi_b \leq 0.5$, Tetlow $et\ al.$ [24] observed shear-induced diffusion phenomena where particles migrate from the high shear-rate region near the inner cylinder toward the low shear-rate region near the outer wall. In the case of a very dilute suspension ($\phi_b = 0.001$), Majji and Morris [1] detected inertial migration, with particles collecting in an equilibrium location near the middle of the annulus with an offset toward the inner cylinder. Our previous work in Kang and Mirbod [6] has also shown the nonuniform particle distribution induced by the shear gradient in the flow, consistent with observations by Tetlow *et al.* [24]. Nevertheless, it is crucial to emphasize that the disparity between our model and the experimental findings of Majji and Morris [1] may stem from the limitations of the SBM, particularly its inability to predict the lift force resulting from both inertia and particle rotation. This underscores the significance of conducting simulations considering all the required forces in the flow and experiments on TC suspensions to more comprehensively understand particle migration in both semi-dilute and concentrated suspensions.

To investigate the nature of the bifurcation for ribbons, we have utilized the Landau model, which expresses the evolution of the flow perturbation in its weakly nonlinear regime and is given by [6,25,26]

$$\frac{dA}{dt} = \sigma(1 + ic_1)A - l(1 + ic_2)|A|^2 A + \cdots .  \tag{7}$$

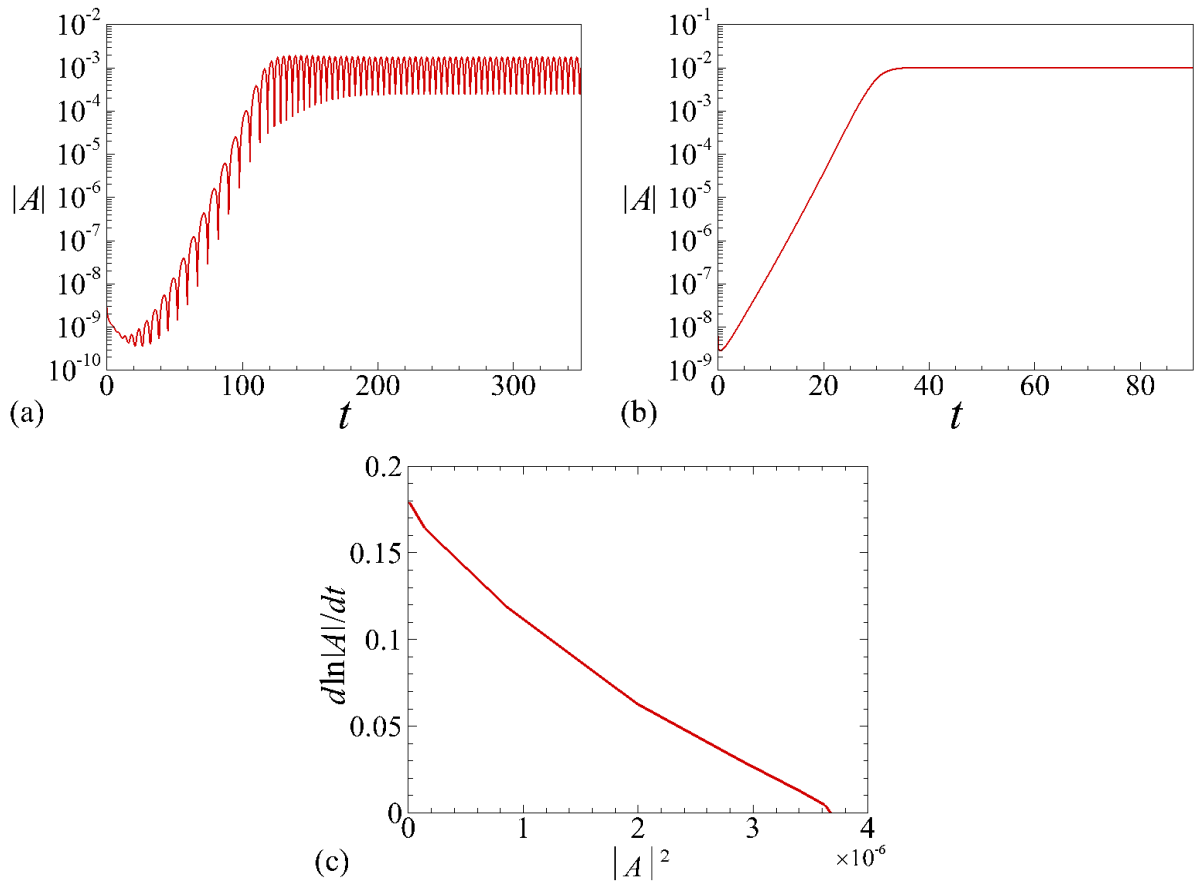

Figure 8. Growth and saturation phases of the perturbation amplitude $|A|$ for (a) $Re_s = 120$ (RIB) and (b) $Re_s = 122$ (SVF) [6] with $\epsilon = 60$ and $\phi_b = 0.1$. (c) The derivative of the amplitude logarithm plotted against the square of the amplitude for $Re_s = 120$, $\epsilon = 60$, and $\phi_b = 0.1$.

The parameters $\sigma$ and $l$ indicate the linear growth rate of the perturbation and the Landau constant. The sign of the constant $l$ determines the nature of the transition (supercritical or subcritical). The constants $c_1$ and $c_2$ are the linear and nonlinear dispersion coefficients, respectively. The norm of the radial velocity component at the central surface in the annulus has been employed to define the amplitude of the perturbation $|A|$ in Eq. (7) as

$$|A| = \frac{1}{2\pi\Gamma}\int_0^\Gamma \int_0^{2\pi} |u_r(r_m, \varphi, z)|\mathrm{d}\varphi\mathrm{d}z, \quad \text{where } r_m = (r_i + r_o)/2. \tag{8}$$

Figures 8(a) and 8(b) show the time history of the amplitude $|A|$ for $Re_s = 120$, where the flow pattern is the RIB state, and for $Re_s = 122$, in which the spiral vortex flow (SVF) occurs in the annulus (see Kang and Mirbod [6]), respectively. At the initial stage of the simulation, a base flow with a simple shear flow is imposed on the flow field with a uniform particle volume fraction ($\phi_b = 0.1$), and low-amplitude random noise of $O(10^{-6})$ is combined in the flow. After the decay of the noise, the bifurcation is triggered, and the amplitude grows exponentially, yielding the growth rate $\sigma$ as the slope of the linear portion of the curve. However, the modes of instability are different for the two flow states. For the case of RIB, the perturbation slowly rises with an oscillation and finally pulsates with constant amplitude (Fig. 8(a)). The oscillation (or oscillatory mode [27,28]) arises from the standing wave of the increasing perturbation, which eventually yields the ribbon pattern in the annulus. On the other hand, the perturbation grows rapidly without an oscillation (stationary mode [27,28]) during the onset of SVF and is saturated with a constant value after the nonlinearity sets in (Fig. 8(b)). We also observed the stationary mode during the occurrence of WSVF and WVF.

As mentioned above, the Landau constant $l$ in Eq. (7) gives the type of transition. The transition is supercritical if $l > 0$, while it is subcritical when $l < 0$ [6,29,30]. The sign of $l$ can be determined from the plot of $\mathrm{d}\ln|A|/\mathrm{d}t$ versus $|A|^2$. The constant $l$ is also given by the slope at the origin ($|A|^2 = 0$). In Fig. 8(c), the instantaneous growth rate $\mathrm{d}\ln|A|/\mathrm{d}t$ against $|A|^2$ is illustrated for $Re_s = 120$ based on the variation of $|A|$ described in Fig. 8(a). The plot reveals that the transition to RIB occurs through a supercritical bifurcation ($l > 0$). As an aside, we note RIB also bifurcates supercritically from the base flow in viscoelastic TCF, as indicated in the theoretical analysis using the Oldroyd-B model [31,32].

In Fig. 9, the modified maps (with different protocols) of flow patterns for a suspension of $\phi_b = 0.1$ and $\epsilon = 60$ have been represented together with the flow patterns of a pure Newtonian fluid. It should be noted that the results were produced with the initial conditions which are circular Couette flow and a uniform particle volume fraction ($\phi_b = 0.1$) for the concentration field. Experimentally, the RIB state appears between CCF and SVF, as observed by Majji *et al.* [2] and Baroudi *et al.* [5] with the same radius ratio. However, RIB eemerges in a narrow range of $Re_s$ compared to the other patterns. The threshold for RIB has been predicted as $115 < Re_s < 120$. This is higher than the experiments [2,5]. Majji *et al.* [2] found the threshold at $Re_s \cong 107$, while the ribbons occurred at $Re_s \cong 110$ in the experiment of Baroudi *et al.* [5].

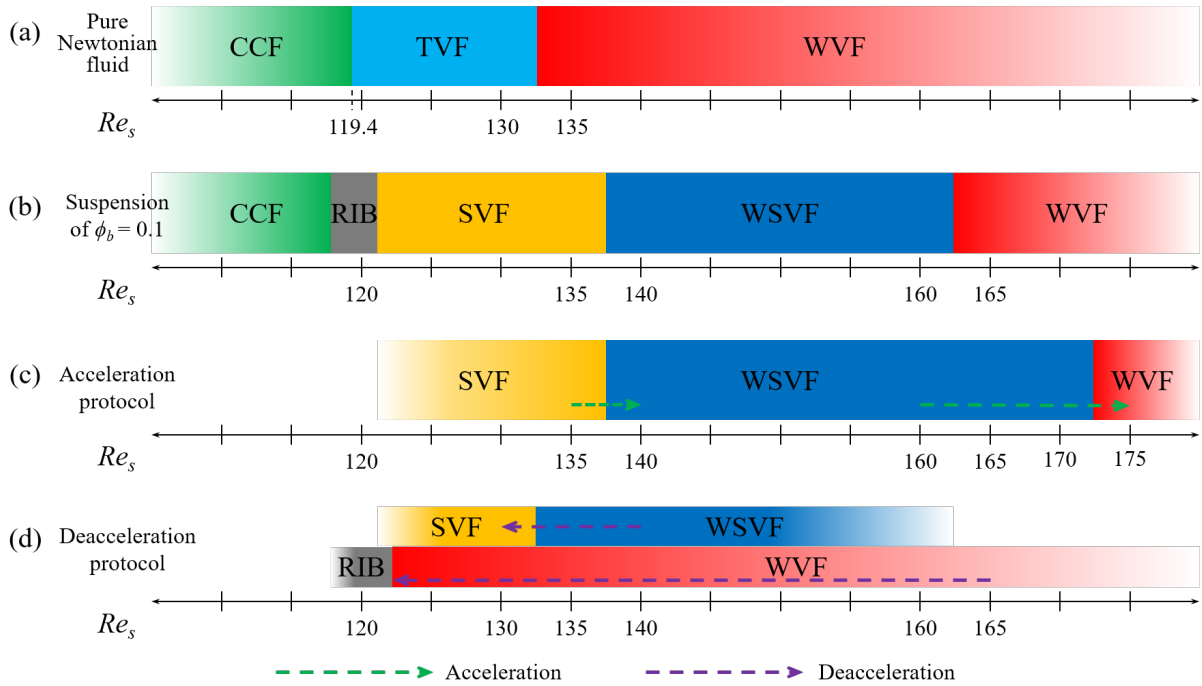

Figure 9. Modified phase diagrams of flow patterns for a suspension of $\phi_b = 0.1$ with $\epsilon = 60$ compared with the diagram for a pure Newtonian fluid (a). In (b), all computations were performed by imposing a simple shear flow with a uniform particle volume fraction on the suspension in the beginning. Phase diagrams with (c) acceleration protocol and (d) deacceleration protocol in a step change.

## Ⅳ. Conclusions

This study investigates the flow of a suspension with concentration $\phi_b = 0.1$ between rotating concentric cylinders using the Suspension Balance Model (SBM) by Nott and Brady [13] and Morris and Brady [14] to model particle stresses. While SBM effectively describes particle stresses when the particle scale Reynolds number is nearly zero, it fails to predict inertial migration observed in experiments by Majji *et al.* [2], Ramesh *et al.* [3], and Moazzen *et al.* [7], where particles with finite inertia undergo non-uniform distribution due to flow curvature in rotating systems.

In our approach, we assumed $Re_p \ll 1$ (strong fluid-particle interactions and negligible particle inertia). Experimental particle Reynolds numbers were also $Re_p=O(10^{-6})$, $O(10^{-5})$, and $O(10^{-2})$ for Majji et al. [2], Ramesh et al. [3], and Baroudi et al. [5], respectively. Despite differences in dominant physics (shear-induced migration in simulations vs. inertial migration in experiments), our studies align well with Majji et al. [2] and Baroudi et al. [5], identifying (1) ribbons as the primary instability and (2) observing a secondary instability in the form of a spiral vortex flow.

We observed hysteresis in secondary bifurcations, specifically in flow transitions occurring during both increasing-$Re_s$ and decreasing-$Re_s$ procedures with step changes in $Re_s$. The transition SVF↔WSVF occurs at $Re_s$ values that differ between ramp-up and ramp-down protocols. Moreover, for the transition WSVF↔WVF, distinct sequences for ramp-up and ramp-down protocols were found. When $Re_s$ is step-increased into the WSVF regime, the suspension flow becomes WVF at $Re_s = 175$. On the other hand, with decreasing $Re_s$ the WVF pattern persists for a wide range of Reynolds numbers well below $Re_s = 175$ and, eventually, the flow transitions to RIB. Accordingly, this behavior clearly illustrates hysteresis of the bifurcations.

Additional simulations have been conducted near the threshold of the instability to investigate the onset of RIB. We have found that RIB can occur via a supercritical bifurcation from CCF; at the onset of instability, and RIB grows very slowly as a standing wave in space, oscillating in time with vortices. As the Reynolds number is further increased, we have found the following transition sequence RIB → SVF → WSVF → WVF for the parameter values examined in this study, similar the sequence observed in experiments (CCF → RIB → SVF → TVF → WVF) In the RIB state, particles are transferred along the convective flow generated by the vortices and undergo shear-induced particle migration caused by the gradient in the shear. Consequently, the particles migrate mostly to the outer cylinder and partly accumulate between counterrotating vortices in the outward flow. The specific distribution of particles for the RIB arises from the shear gradient in the radial direction which is larger than the gradient in the vortices, whereas it is stronger near the core of vortices in the TVF and SVF states. Accordingly, the region of higher concentration appears when the suspension flows toward the outer wall, as the outward flow weakens the shear gradient in the radial direction and thus it limits the shear-induced migration. Future research endeavors can concentrate on comprehensively characterizing and explaining the hysteresis phenomenon, determining if analogous or novel patterns emerge during both the ramp-up and ramp-down processes across different suspension concentrations. Additionally, conducting experiments in a suspension TC cell with both narrow and wide gap configurations would be intriguing. This approach aims to unravel the influence of particles and their migration behavior, especially concerning diverse particle volume fractions within the annulus region.

## Acknowledgments

This work was supported by the Basic Science Research Program through the National Research Foundation of Korea (NRF) funded by the Ministry of Education (NRF-2021R1I1A3048306) to C.K. We also acknowledge NSF award no. 1854376 and ARO award no. W911NF-18-1-0356 to P.M.

## Declaration of interests

The authors report no conflict of interest.

https://doi.org/10.1098/rspa.1994.0132